\documentclass[aip,jcp,reprint]{revtex4-1}

\usepackage{amsmath}
\usepackage{graphicx}% Include figure files
\usepackage{color}
\usepackage{dcolumn}% Align table columns on decimal point
\usepackage{bm}% bold math
\usepackage{textcomp}
\usepackage{natbib}

\begin{document}

\title{Optical Kerr effect of liquid and supercooled water: the experimental and data analysis perspective.}

\author{A. Taschin$^{1}$, P. Bartolini$^{1}$, R. Eramo$^{1,2}$, R. Righini$^{1,3}$, and R. Torre$^{1,4}$}

\affiliation{
$^1$European lab. for Non-Linear Spectroscopy (LENS), Univ. di Firenze, via N. Carrara 1, I-50019 Sesto Fiorentino, Firenze, Italy.\\
$^2$Istituto Nazionale Ottica, CNR, Largo Fermi 6, I-50125 Firenze, Italy.\\
$^3$Dip. di Chimica, Univ. di Firenze, via Della Lastruccia 13, I-50019 Sesto Fiorentino, Firenze, Italy.\\
$^4$Dip. di Fisica e Astronomia, Univ. di Firenze, via Sansone 1, I-50019 Sesto Fiorentino, Firenze, Italy.}

%\preprint{version 00}
\date{\today}

\email[Corresponding Author: R.Torre, ]{torre@lens.unifi.it}

\begin{abstract} 

The time-resolved optical Kerr effect spectroscopy (OKE) is a powerful experimental tool enabling accurate investigations of the dynamic phenomena in molecular liquids. We introduced innovative experimental and fitting procedures, that permit a safe deconvolution of sample response function from the instrumental function. This is a critical issue in order to measure the dynamics of sample presenting weak signal, e.g. liquid water. We report OKE data on water measuring intermolecular vibrations and the structural relaxation processes in an extended temperature range, inclusive of the supercooled states. The unpreceded data quality makes possible a solid comparison with few theoretical models; the multi-mode Brownian oscillator model, the Kubo's discrete random jump model and the schematic mode-coupling model. All these models produce reasonable good fits of the OKE data of stable liquid water, i.e. over the freezing point. The features of water dynamics in the OKE data becomes unambiguous only at lower temperatures, i.e. for water in the metastable supercooled phase. Hence this data enable a valid comparison between the model fits. We found that the schematic mode-coupling model provides the more rigorous and complete model for water dynamics, even if is intrinsic hydrodynamic approach hide the molecular informations.  

\end{abstract}
\maketitle

\section{Introduction}

Heterodyne-detected optical Kerr effect (HD-OKE) has been widely used since more than twenty years for the investigation of dynamical properties of molecular liquids. Not surprising, a remarkable fraction of this work has been dedicated to the study of liquid water, often in relation to the hypothesized presence of two liquid phases. We recently published new HD-OKE experimental results covering a very broad temperature range, extended into the supercooled regime, characterized by very high accuracy and unprecedented signal-to-noise ratio \citep{taschin_13}. Here we intend to report on the detailed analysis of those data, comparing the results of different theoretical models. Our intention is not to formulate a ranking of those models on the basis of their ability of reproducing the experimental data, but rather to highlight the pros and cons of the different approaches and, most of all, to point out the key features of each model responsible for its predictive capabilities. 
In the following, we first recall the main aspects of HD-OKE measurements of liquid and supercooled water, and of the recovery of the sample response function, both in time and frequency domains. We dedicate special attention to the deconvolution of the femtosecond OKE data, and demonstrate the crucial importance of an accurate measurement of the instrumental function. In the second part, we present the detailed analysis of those data making use of three different theoretical models. We conclude by summarizing the main findings and pointing out the most relevant features of the models.

\section{HD-OKE measurements of liquid water}

In a HD-OKE experiment~\cite{righini_93,hunt_07,bartolini_08} a linearly polarized laser pulse induces transient birefringence in a medium by means of a non resonant non-linear effect. The induced birefringence can be probed by a second pulse of different polarization, spatially superimposed to the pump within the sample. The change of the polarization status of the probe pulse, measured as a function of the delay time from the pump, gives information about the non-linear response of the studied material. The signal is characterized by an instantaneous electronic contribution plus a decaying contribution, which constitutes the most interesting part, as it contains information about the relaxation and vibrational response of the molecules in the sample.
The time window probed in an OKE experiment can be very broad, extending from tens of femtoseconds to hundreds of picoseconds. This makes OKE a very powerful technique, capable of revealing very different dynamic regimes in a variety of samples, from simple liquids to supercooled liquids and glass formers~\cite{bartolini_99,torre_98,torre_00,prevosto_02,ricci_02,ricci_04}, it makes possible measuring, at the same time, the slow relaxation processes and the fast inter molecular vibrations of the system. In the optical heterodyne detection configuration, the signal is directly proportional to the third order non-linear response function of the material, $R(t)$, convoluted with the instrumental function, $G(t)$.

\begin{figure*}[t]
\includegraphics[scale=0.6]{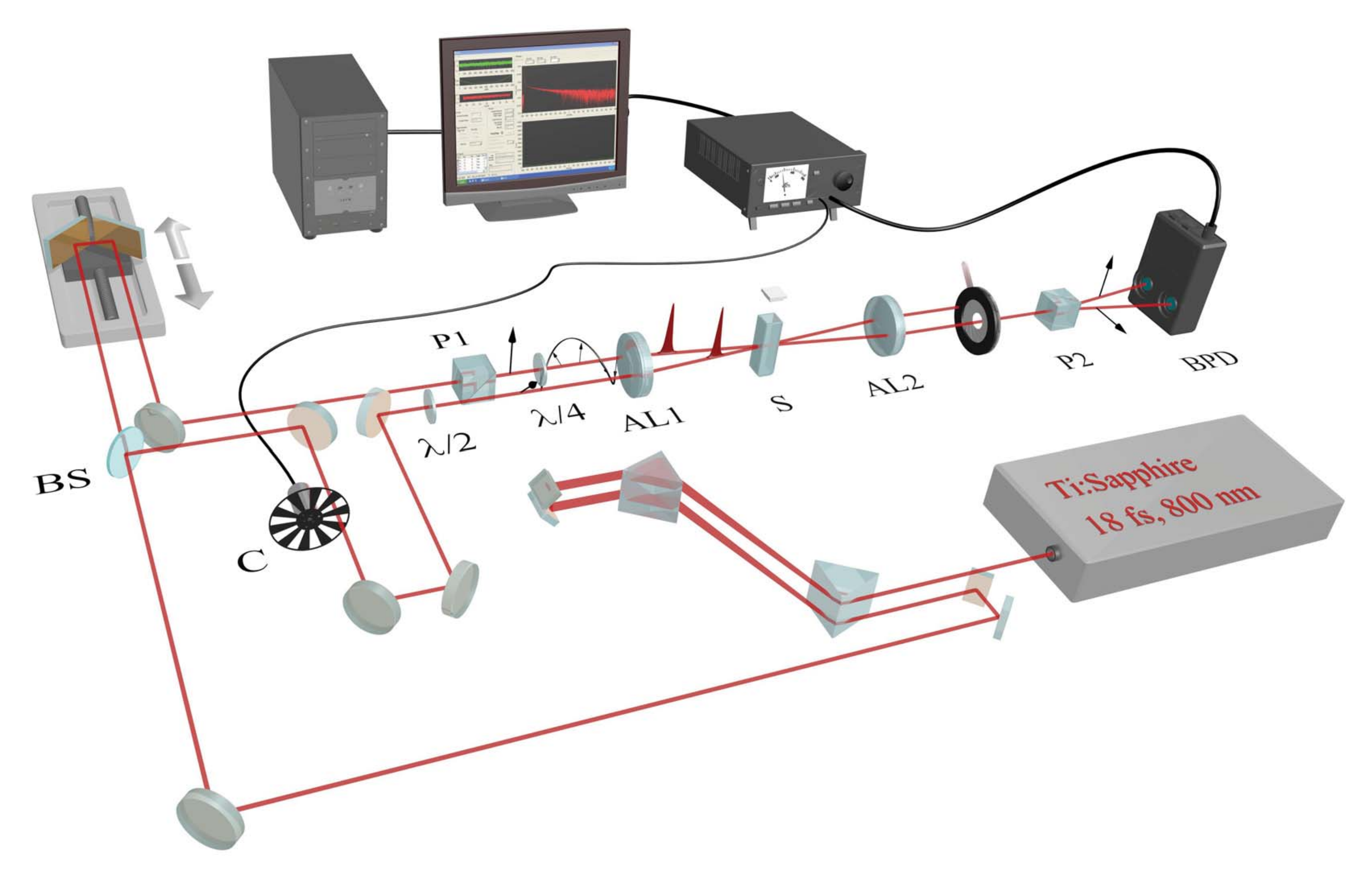}
\caption{
The optical set-up for heterodyne-detected optical Kerr effect measurements (HD-OKE). The laser pulses are produced by a Ti:Sapphire Kerr-lens mode locked cavity and their group velocity dispersion is controlled by a prism compression stage. The laser beam is split by a beam splitter (BS) in the pump and probe beams. The probe pulse is delayed respect to the pump pulse by an computer controlled optical delay line. A half wave plate ($\lambda/2$) fixes the polarization of the pump at $45^\circ$ from that of the probe, set vertical by the P1 polarizer. The probe polarization is then converted to circular by a quarter wave plate ($\lambda/4$). Probe and pump beams are focused inside the sample S by the achromatic lens AL1. The probe beam is then re-collimated by the achromatic lens AL2 and sent to the Wollaston polariser P2. The horizontal and vertical linear polarizations, both present in the probe, are selected by P2. 
This optical set-up produces twin OKE signals with vertical and horizontal polarizations having opposite phase with the local field. The balanced photodiode detector BPD operates the electronic subtraction of the two polarization components and extracts the HD-OKE signal from the background. The analogue output of the photodiode is sent to a lock-in amplifier together with the reference signal of chopper (C) placed on the pump beam. A DAQ board simultaneously acquires the translation stage position from the encoder board and the lock-in output signal. Finally, the signal and position are stored by the computer to build the final signal time profile.}
\label{setup}
\end{figure*}

Liquid water is a challenging sample for this kind of experiment because of its very weak OKE response, due to its nearly isotropic molecular polarizability. 
For this reason, we implemented in our set-up, based on a Ti:Sapphire laser oscillator (wavelength 800 nm, pulse width 18 fs), two experimental improvements that enabled us to measure the fast vibrational dynamics and the slow structural relaxation in the same experiment, with very high signal-to-noise ratio and large dynamic range. The first feature is the independent and continuous motion of a translation stage, equipped with a linear encoder, which ensures the absolute control of the position~\cite{bartolini_07}. This reduces substantially the acquisition time and improves the signal statistics. The second one is the implementation of a peculiar configuration of the heterodyne detection~\citep{giraud_03,bartolini_09}, which makes use of a circularly polarized probe beam and of a differential detection of two opposite-phase signals on a balanced photodiode. As shown in fig.\ref{setup}, a quarter wave-plate between the two polarisers produces the circularly polarized probe field. Two signals, with opposite polarizations and with opposite phase respect to the local field, emerge from the P2 polarizer, and are sent to the balanced photodiode detector. The OKE signal measured in such a way is automatically heterodyned and free from any spurious phase-independent signal. A further improvement is obtained by subtracting the two HD-OKE measurements obtained with left and right circular polarizations of the probe field. This procedure removes from the signal the dichroic contributions coming from possible misalignment of the wave-plate. The output signal of the photodiode is amplified by a lock-in amplifier and digitalized by an acquisition board. A home made software acquires the processed signal together with the reading of the delay line encoder and retraces the final time dependent HD-OKE signal.

Supercooling bulk water is not an easy task because water is prone to crystallization; special care in the preparation and manipulation of the samples is required in order to reach very low temperatures. We performed our measurements on a sealed vial of cylindrical shape, prepared for pharmaceutical purposes by the Angelini company: the lowest temperature reached for this sample was 247 K. The vial was inserted into a parallelepiped-shaped aluminum holder, whose central cylindrical cavity fits the diameter of the vial. A thin film of glycerol between the vial and the housing assured an efficient heat transfer. The holder was fixed to the cold plate of a Peltier cooler, whose temperature was controlled, with a stability of 0.1 K, by a platinum thermoresistance in thermal contact with the holder itself. Two fused silica windows inserted on two opposite sides of the aluminum holder allowed the beams to cross the sample. 

\subsection{\label{sec:OKEsignal}HD-OKE signal}

The signal measured in an heterodyne-detected optical Kerr effect (HD-OKE) experiment is \cite{torre_99,prevosto_02,bartolini_08,bartolini_09,kinoshita_12}
\begin{equation} \label{signal1}
\begin{split}
	S(\tau) & \propto \int_{-\infty}^{+\infty} dt I_{pr}(t-\tau) \int_{-\infty}^{+\infty} dt' R(t-t') I_{ex}(t') \\
	& = \int_{-\infty}^{+\infty}dt_1 R(t_1)G(\tau - t_1)
\end{split}
\end{equation}
with
\begin{equation}
G(t)=\int_{-\infty}^{+\infty}dt_2 I_{pr}(t_2) I_{ex}(t_2+t)
\label{crosscor}
\end{equation}
where $I_{pr}$ and $I_{ex}$ are the probing and exciting laser intensities, respectively; $G(t)$ is their intensity correlation, determining the experimental time resolution and has thus the role of instrumental function, and $R(t)$ is the material response. 

Since we are performing a non-resonant OKE experiment, the Born-Oppenheimer approximation applies and the response function can be cast in the form~\cite{hellwarth_77,foggi_92,torre_93,ricci_93,ricci_95,torre_98b}:
\begin{equation}
R(t)=\gamma\delta(t)+R_n(t)
\label{signal2}
\end{equation}
with $\gamma$ representing the instantaneous electronic response and $R_n(t)$ the nuclear response. The latter can be written in the classical limit as
\begin{equation}
R_n(t)\propto-\frac{\theta(t)}{kT}\frac{\partial}{\partial t}\Phi_{\chi\chi}
\label{Response}
\end{equation}
In eq. \ref{Response} $\theta(t)$ is the Heaviside step function, $k$ is the Boltzmann constant and $\Phi_{\chi\chi}$ the time correlation function of the anisotropic susceptibility
\begin{equation}
\Phi_{\chi\chi}=\langle\chi_{xy}(t)\chi_{xy}(0)\rangle
\label{CorrelFunction}
\end{equation}
$\chi_{xy}(t)$ beaing the off-diagonal element of the susceptibility tensor (i.e. the collective electronic polarizability). 

In order to fit the OKE data we simulated the measured signal using the following expression:
\begin{equation}
S(t)\propto\int_{-\infty}^{+\infty} \left[ \gamma\delta(t-t')+R_n(t-t') \right]  G(t') dt'
\label{signalfit}
\end{equation}
Taking the Fourier transform of eq. \ref{signalfit}, we get
\begin{equation}
\widetilde{S}(\omega)=\left[ \gamma + \widetilde{R}_n(\omega)\right] \widetilde{G}(\omega)
\label{signalfitFT}
\end{equation}
as, for the non resonant OKE, $\gamma$ is real, we get from \ref{signalfitFT} the important result that the imaginary part of $\widetilde{S}(\omega)$ is unaffected by the instantaneous response, and that for nuclear part
\begin{equation}
Im\left[ \widetilde{R}(\omega)\right] =Im\left[ \frac{\widetilde{S}(\omega)}{\widetilde{G}(\omega)}\right]
\label{imromega}
\end{equation}
hallowing the  extraction of $\widetilde{R}_n(\omega)$ from the OKE signal once the instrumental function is known.

\subsection{Instrumental function measurement}
\label{InstrFunct}

An important experimental issue concerns the measurement of the actual instrumental function $G(t)$, which is far from trivial. Eq.\ref{signalfit} shows clearly that the instrumental function can be measured directly if the sample has negligible or very fast nuclear response (i.e. if $R_n(t)\propto\delta(t)$). 

In many cases, a fused silica plate has been used to this purpose, because of its weak and fast nuclear response. Actually, the OKE response of silica is quite complex and difficult to determine, so it turns out to be not particularly adapt when a precise determination of real $G(t)$ is required. 
Some previous studies utilized the second harmonic cross-correlation of the pump and probe pulses to measure the instrumental function, see Kinoshita et al.~\cite{kinoshita_95}. We found that any small modification and adjustment of the experimental set-up, inevitable when replacing the water sample with the reference material chosen for measuring $G(t)$, causes severe alterations of the latter. For instance, i) the insertion of a second harmonic crystal in place of the sample modifies the pulse compression status, ii) the spatial overlap the beams requires some re-alignment, iii) the translation stage has to be re-positioned in order to achieve the temporal superposition of pump and probe pulses. Actually, even small changes of the experimental conditions critically affect the instrumental function. 
With an alternative procedure, in some other cases the instrumental function was obtained by fitting the instantaneous electronic contribution to the HD-OKE signal with an analytic peak function (Gaussian, hyperbolic secant, etc..).

The OKE experiments in water are typically characterized by very low
and fast signals, so that the precise determination of the instrumental function is extremely important in order to extract the water response. The experimental methods, summarized above, are not enough accurate for this purpose. 

We measured the instrumental function following a different procedure that grants the proper level of accuracy required for water investigations.  As reference sample we chose a plate of calcium fluoride (CaF$_2$); this is a cubic ionic crystal with only one Raman active band in the probed frequency range, the optical phonon at 322~cm$^{-1}$ of T$_{2g}$ symmetry. Its nuclear OKE response is then simple and well known. The calcium fluoride plate was dipped inside a water vial, identical to that used for water measurements, supported by the same sample holder. The measurement of the instrumental function was done by just replacing the water vial with the water-CaF$_2$ vial, leaving the rest of the set-up unchanged. We took care that the faces of the CaF$_2$ plate were perpendicular to the line bisecting the angle formed by pump and probe beams. The thickness of the CaF$_2$ plate, 3 mm, was enough to fully contain the probe and pump overlap area, avoiding any spurious signal contribution by the outer water. We took a reference measurement for each set of water data. 

\begin{figure}[htb]
\includegraphics[scale=0.5]{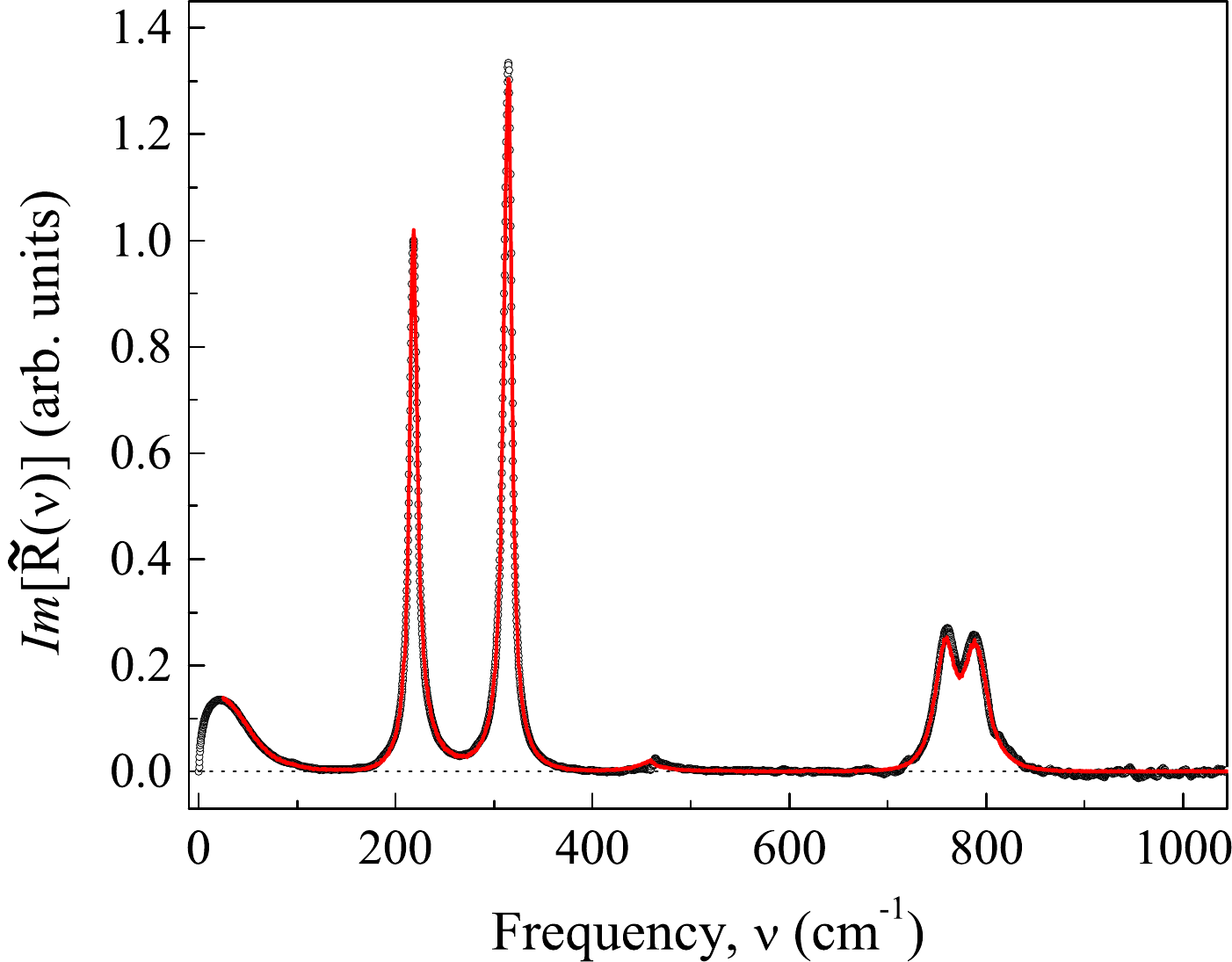} 
\caption{Frequency response of CCl$_{4}$ obtained from the OKE data using the instrumental function extracted from a CaF$_2$ slab inserted in the CCl$_{4}$ cell. Our data (black open circles) reproduce very well the depolarized Raman scattering spectrum (red continuous line), both in frequency and amplitude, corrected for the Bose factor.}
\label{CCl4-freq}
\end{figure}

This procedure, differently from the other approaches, allows us to accurately preserve the experimental conditions adopted in the water and in the reference measurements.  
%
%figure 2
\begin{figure}[htb]
\includegraphics[scale=0.5]{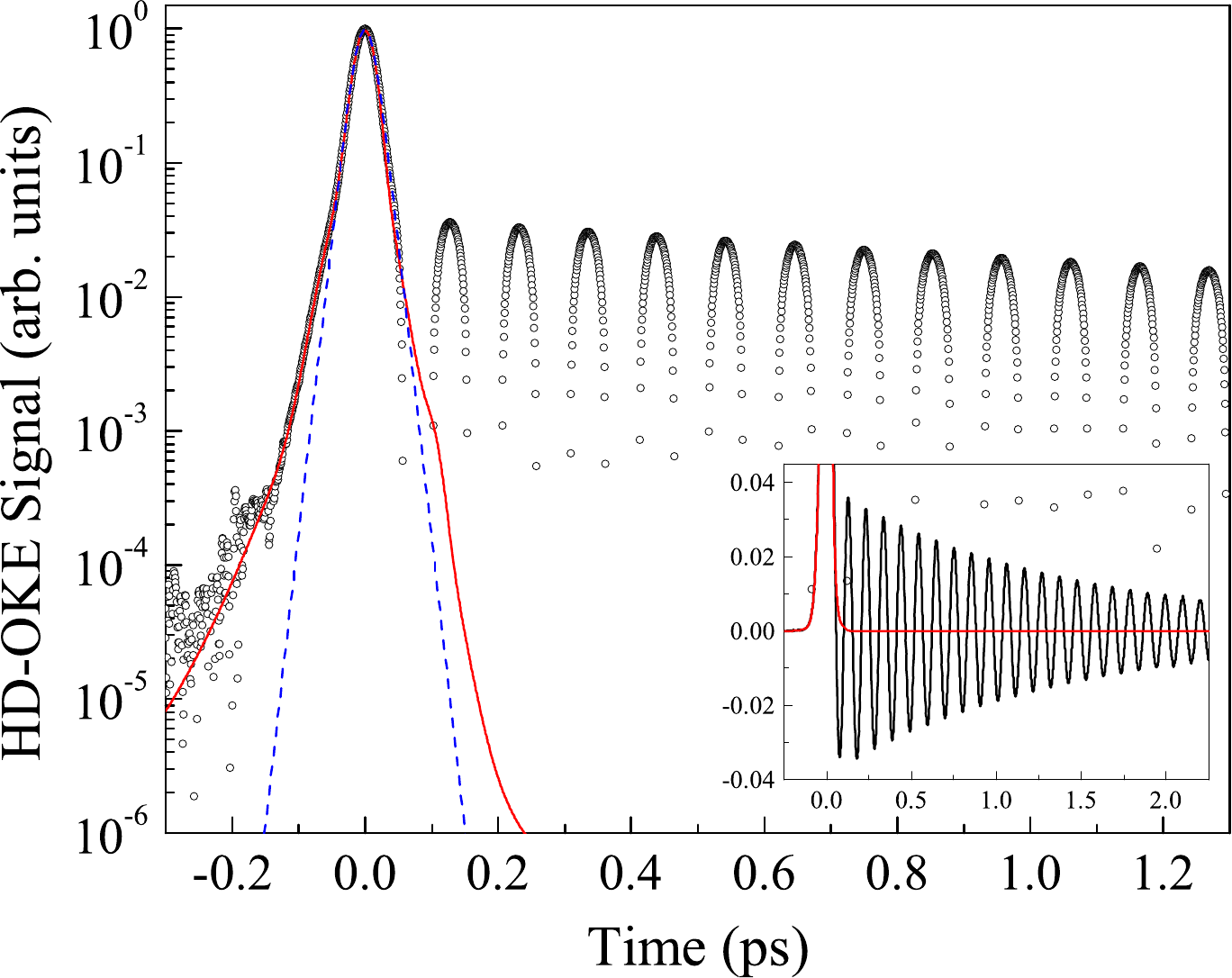}
\caption{A typical HD-OKE signal of the reference sample and the extracted instrumental function. In the figure we show the HD-OKE signal of CaF$_2$ (circles) and the instrumental function (red continuous line) obtained with the fitting procedure described in the text. The nuclear response, $R_n$, is taken as the time-derivative of a single damped oscillator and the instrumental function, $G$, is simulated as the sum of Gaussian, Lorentzian, and hyperbolic secant functions. The comparison with the fit of the instantaneous electronic part performed using just a single hyperbolic secant (blue dashed line) is reported too.}
\label{CaF2}
\end{figure}
Fig.~\ref{CaF2} reports a typical HD-OKE signal obtained in the CaF$_2$ reference sample: the signal shows a first peak, due to the electronic response, characterized by very fast rise and fall, and a second oscillating contribution due to the nuclear response. This signal is the convolution of the instrumental function with the OKE response, see eq.\ref{signalfit}; for CaF$_2$ its nuclear part is the time derivative of a single harmonic damped oscillator (DHO). The simplicity of this nuclear response function allows the reliable extraction of the instrumental function, $G(t)$, by means of an iterative fitting procedure (i.e. least square fitting of the HD-OKE data of the reference sample with the simulated signal according to eq.\ref{signalfit}). We found that our instrumental function can not be reproduced by a single hyperbolic secant function, as often
reported in the literature; a good fit requires the sum of several functions, namely a combination of Gaussian, Lorentzian and/or hyperbolic secant functions. In Fig.~\ref{CaF2} we report the instrumental function obtained by the iterative fitting procedure (red continuous line) and the one obtained by fitting the electronic peak with a simple hyperbolic secant function (blue dashed line).
Apparently these two instrumental functions are very similar but the small differences cannot be neglected for an accurate investigation of water fast OKE response. This is mostly evident when the data are Fourier transformed to the frequency domain.
%
%figure 3
\begin{figure}[htb]
\includegraphics[scale=0.5]{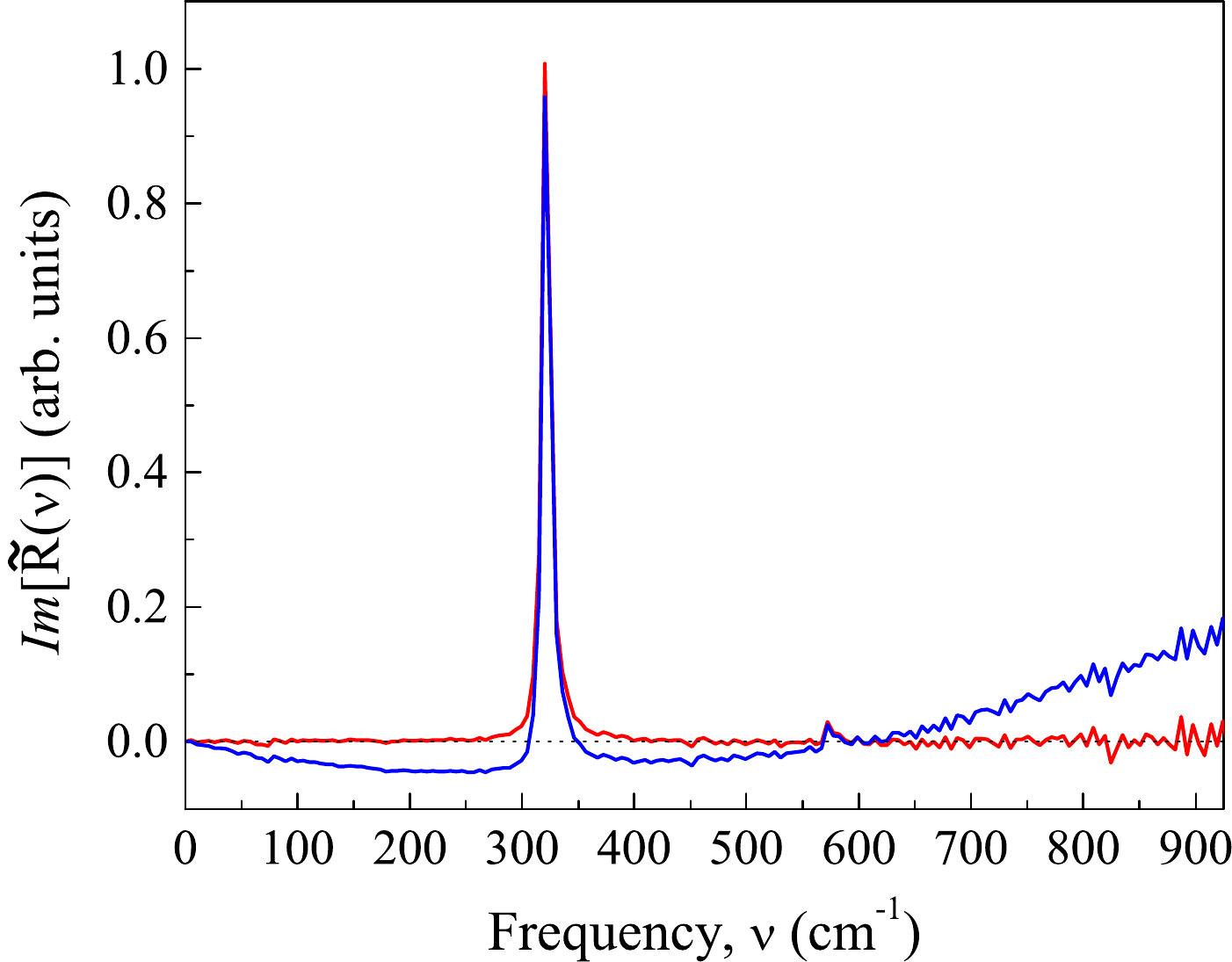} 
\caption{Dependence of the reference sample spectrum on the choise of the instrumental function $G(t)$. We obtained the response function by deconvolution of the HD-OKE signal $S(t)$ according to eq. \ref{imromega}.
The figure compares the results obtained for the CaF$_2$ sample by adopting the iterative fitting procedure described in the text (red line), and by fitting the electronic peak with a hyperbolic secant function (blue line). Only the knowledge of the “real” instrumental function allows measuring the correct response.}
\label{CaF2-freq}
\end{figure}
Fig.~\ref{CaF2-freq} shows the comparison between the frequency response calculated using the instrumental function obtained by the iterative fitting procedure (red line) and the one obtained by simply fitting the electronic peak with a hyperbolic secant function (blue line). We clearly see that only an accurate measurement of the real time domain instrumental function yields the correct response. 
As a further test of the method, we measured the OKE signal of a carbon tetrachloride (CCl$_{4}$) liquid sample. We inserted a scrap of CaF$_2$ slab in the  CCl$_{4}$ cell for measuring the instrumental function. The resulting CCl$_{4}$ frequency response is shown in fig.~\ref{CCl4-freq} (black open circle). The agreement of our data with the depolarized Raman scattering spectrum (red continuous line), corrected for the Bose factor, is good in the entire spectral range; frequencies and amplitudes of Raman bands being very well reproduced. This confirms the high accuracy of the employed method.
%figure4

The role of the instrumental function in HD-OKE experiments has been always considered as a underpinning problem for a correct data analysis, but to our knowledge it has never been addressed in such a quantitative way. In the present study, the unprecedented signal/noise ratio and the accurate determination of the instrumental function ensure a correct data analysis, from which a reliable OKE response can be extracted, also in case of weak and fast decaying signals.

\subsection{HD-OKE water data}
%figure5
\begin{figure*}[t]
\includegraphics[scale=0.9]{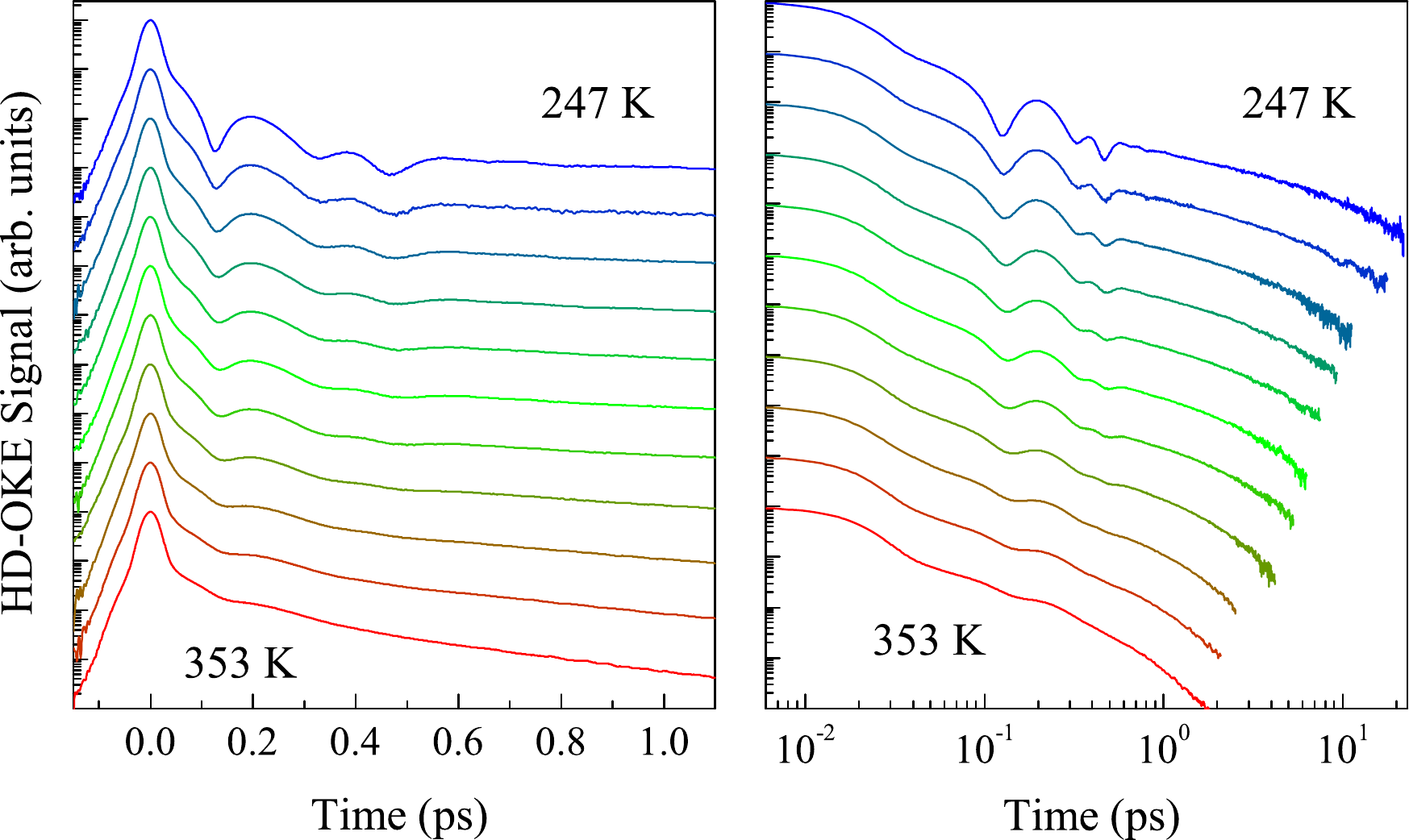} 
\caption{Log-linear (left panel) and log-log (right panel) plots of the HD-OKE data on liquid and supercooled water. From the bottom to the top the following temperatures are reported: 353, 333, 313, 293, 278, 273, 268, 263, 258, 253, $247~K$. : the oscillations due to the inter-molecular vibrational modes, clearly visible at short times (left panel) smoothly merge at longer times into the structural relaxation decay (see right panel). By lowering the temperature, the vibrational dynamics becomes more defined and the monotonic decay longer and strongly non-exponential.}
\label{OKE-data}
\end{figure*}
In Fig.~\ref{OKE-data} we report all the HD-OKE data collected on water at changing of temperatures from liquid phase to the supercooled one. The left panel shows the short times in a log-linear plot and the right panel the data are reported in a log-log plot showing the whole time-scale measured. Water shows a complex relaxation pattern strongly dependent by the temperature. Decreasing temperature the fast oscillating dynamics becomes more structured and the slow relaxation decay becomes increasingly long. 

At short times we have fast oscillations due to inter-molecular vibrational modes. These correspond to the two broad bands centered at about $50~cm^{-1}$ and $200~cm^{-1}$, generally addressed in the literature~\cite{walrafen_86,skaf_05,desantis_04,padro_04} as ``bending" and ``stretching" modes of the hydrogen-bond network, respectively. 
At long times the signal shows a monotonic decay; in the first OKE investigations it was interpreted as a bi-exponential relaxation, due to single molecule orientational dynamics\cite{winkler_00}. Further experiments, extending the temperature range to the supercooled phase, proved that the slow decay follows a stretched exponential function, typical of structural relaxation phenomena, with a critical slow down of the relaxation times\citep{torre_04}.

\section{Data analysis and OKE response models}
 
As we briefly summarized in section Sec.\ref{sec:OKEsignal}, the time-resolved HD-OKE experiment measures the time correlation function of the off-diagonal susceptibility elements. These are collective polarizability tensors of the liquid whose definition is quite complex. There have been many studies concerning the basic problem of defining the optical observables (i.e. susceptibility tensor) starting from the molecular features and dynamics, see for example ref.~\cite{hellwarth_70,hellwarth_77,berne_76,balucani_94,torre_08} and references therein. The general theories that define rigorously this connection necessarily involve a huge number of physical variables and they turn out to be not operative for a comparison with the OKE experimental data. So the OKE response interpretation have been typically done using phenomenological models\cite{torre_93,ricci_95,torre_95,torre_96,torre_98b,bartolini_99,bartolini_08},  and/or computer simulations\cite{paolantoni_02,ryu_04,tao_06}. Recently, mode-coupling theories, dynamic models at the mesoscopic scale, based on the memory function approach, have also been used to interpret the OKE results~\cite{torre_98,torre_99,torre_00,prevosto_02b,prevosto_02,ricci_02,pratesi_03,ricci_04,bartolini_08}.

The definition of the OKE response function in liquid water is even more complex than in other liquids due the almost isotropic molecular polarizability and to the hydrogen-bounded network that make the OKE observable dominated by the collective susceptibility and dynamics. 
Numerical simulations of the OKE signal in liquid water\cite{skaf_05,sonoda_05,lupi_12} show a complex interplay
between intrinsic molecular terms and interaction-induced contributions. 
Two main issues need to be addressed in order to define the OKE response. First, one has to pinpoint the physical parameters relevant for the experimentally probed dynamics and what are the equations of motion that they follow. In order to have an operative model, both the liquid modes and their equations of motion must be relatively simple, in other words they should be defined with a coarse-graining approach, where the fine molecular features are averaged out. Second, the connection between optical susceptibility and modes of the liquid must be defined.

Few phenomenological models have been utilized to analyse the whole time-dependent OKE response in liquid water\cite{palese_96,winkler_02,ratajska_06}. As well as, the mode-coupling theory has been used to simulate the water response\cite{torre_04,bartolini_08,taschin_13}.  

In the following sections we compare the results of three different theoretical models in the an analysis of our new OKE data, extending into supercooled water phase; namely, the Multi-mode Brownian Oscillator (MBO) model\cite{tanimura_93,mukamel_95}, the Kubo's Discrete Random Jump model\cite{kubo_62,kubo_69}, and the Schematic Mode-Coupling model~\cite{goetze_92,goetze_00b,goetze_04,goetze_09}. The first two models have been already applied for the analysis of the OKE data of water\cite{palese_96,winkler_02}, but only in the stable liquid phase, the third one has been very recently applied with success to water in the liquid and supercooled phases by the same authors of the present paper~\citep{taschin_13}. Including the HD-OKE results for the supercooled phase of water provides a more stringent test also of the first two theories.

\subsection{Multi-mode Brownian Oscillator Model}
A relatively simple Multi-mode Brownian Oscillator Model (MBO) was utilized by Palese et al. \cite{palese_96} to describe the liquid water dynamics. The model aims at describing the whole relaxation behaviour of the liquid without a time scale separation, a priori imposed, between the fast and slow dynamics.   

The susceptibility tensor is taken as a second order expansion on the nuclear $Q$-modes\cite{tanimura_93}, $\chi(t)\simeq a_1Q(t)+a_2Q^2(t)$. 
The $Q$ variables must be interpreted as local normal modes of the liquid obtained from a coarse-graining treatment of the molecular and intermolecular coordinates. The equation of motion of these MBO modes are described by a Damped Harmonic Oscillator (DHO) equation, $\ddot{Q}(t)-\gamma\dot{Q}(t)+\omega^2 Q(t)=0$. The non-resonant nuclear third-order response function, in particular the OKE response, from a single DHO can be written as\cite{tanimura_93,mukamel_95,palese_96}:

\begin{widetext}
\begin{equation}
R^{MBO}(\Omega,t)=\theta(t)\frac{e^{-\gamma}\frac{t}{2}}{\Omega}\sin(\Omega t)\left\{a_1^2  
+\frac{a_2^2}{\Omega}\left[\coth(\frac{i \hbar}{2kT}\varphi)e^{-\varphi t} - \coth(\frac{i \hbar}{2kT} \bar{\varphi})e^{-\bar{\varphi} t}\right] \right\}
\label{RespMBO}
\end{equation}
\end{widetext}

where $\Omega=\sqrt{\omega^2-\frac{\gamma^2}{4}}$, $\varphi=(\frac{\gamma}{2}+\textit{i}\Omega)$, and $\bar{\varphi}=(\frac{\gamma}{2}-\textit{i}\Omega)$. 
The $a_2$ quadratic term in this equation expresses the non-linear coupling between the susceptibility and the $Q$-mode, when $a_2=0$ the simple DHO response function is recovered. 

The MBO model used by Palese et al. considers a continuous ensemble of $Q$-modes characterized by different frequencies $\omega$; these modes are uncoupled (i.e. defined by independent DHO equations) and all characterized by identical damping coefficients, $\gamma$. Consequently, some modes are under-dumped and other ones are over-damped, in dependence of their $\omega$ value. Each $Q$-mode has a homogeneous broadening, expressed by $\gamma$, which is assumed temperature dependent and related to the liquid viscosity. The collection of $Q$-modes is shaped by an inhomogeneous broadening function, $S(\Omega)$, which fixes the weigh of each mode in the distribution. The inhomogeneous broadening is due to the interactions of the $Q$-mode with the thermal bath and it taken in the form\cite{palese_96}:
\begin{equation}
S(\Omega)=\sum_{n}\frac{\Omega^2 A_n \Gamma_n}{2\pi((\Omega^2-\Omega_n^2)^2 +\Omega^2 \Gamma_n^2)}
\label{SOmega}
\end{equation}
where the sum over $n$ accounts for different dynamics, like librations, translations and bendings, which contribute to the frequency distribution of the harmonic oscillators.
The MBO model simulates the OKE nuclear response function with the oscillator ensemble according to the following integral equation:
\begin{equation}
R_n(t) \propto \int_{\Omega_c}^{\infty} S(\Omega)R^{MBO}(\Omega, t)d\Omega
\label{RespMBOtot}
\end{equation}
The integration range in Eq.~\ref{RespMBOtot} is lower limited by the cut-off frequency $\Omega_c$, corresponding to the physical restriction that no oscillator can have an oscillation period longer than the structural rearrangement of the bath. This cut-off is proportional to the homogeneous damping
\begin{equation}
\Omega_c = b\,\gamma(T),
 \label{OmegaC}
\end{equation}
 where $b$ is a temperature independent coefficient. 

The cut-off frequency in the integration yields a signal that at long times relaxes as a single exponential with time constant $\tau \propto 1/\gamma$. The MBO model cannot account for the stretched exponential decay at very long times; this is a first serious limit in order to fit the OKE data in the supercooled phase.
Moreover, since the structural relaxation time, $\tau$, is proportional to the viscosity, the homogeneous width $\gamma$ is inversely proportional to the viscosity; so the temperature dependence of $\gamma$ is fixed by the viscosity. In summary, in the MBO model the slow relaxation dynamics is accounted for by the superposition of low-frequency over-damped oscillators in eq.\ref{RespMBOtot}, critically dependent on the $\Omega_c$ cut-off frequency, while the fast vibrational part results from the inhomogeneous distribution of under-dumped oscillators. The integral equation (8) provides a smooth merging of the two oscillator ensembles.
%figure6
\begin{figure*}[t]
\includegraphics[scale=0.7]{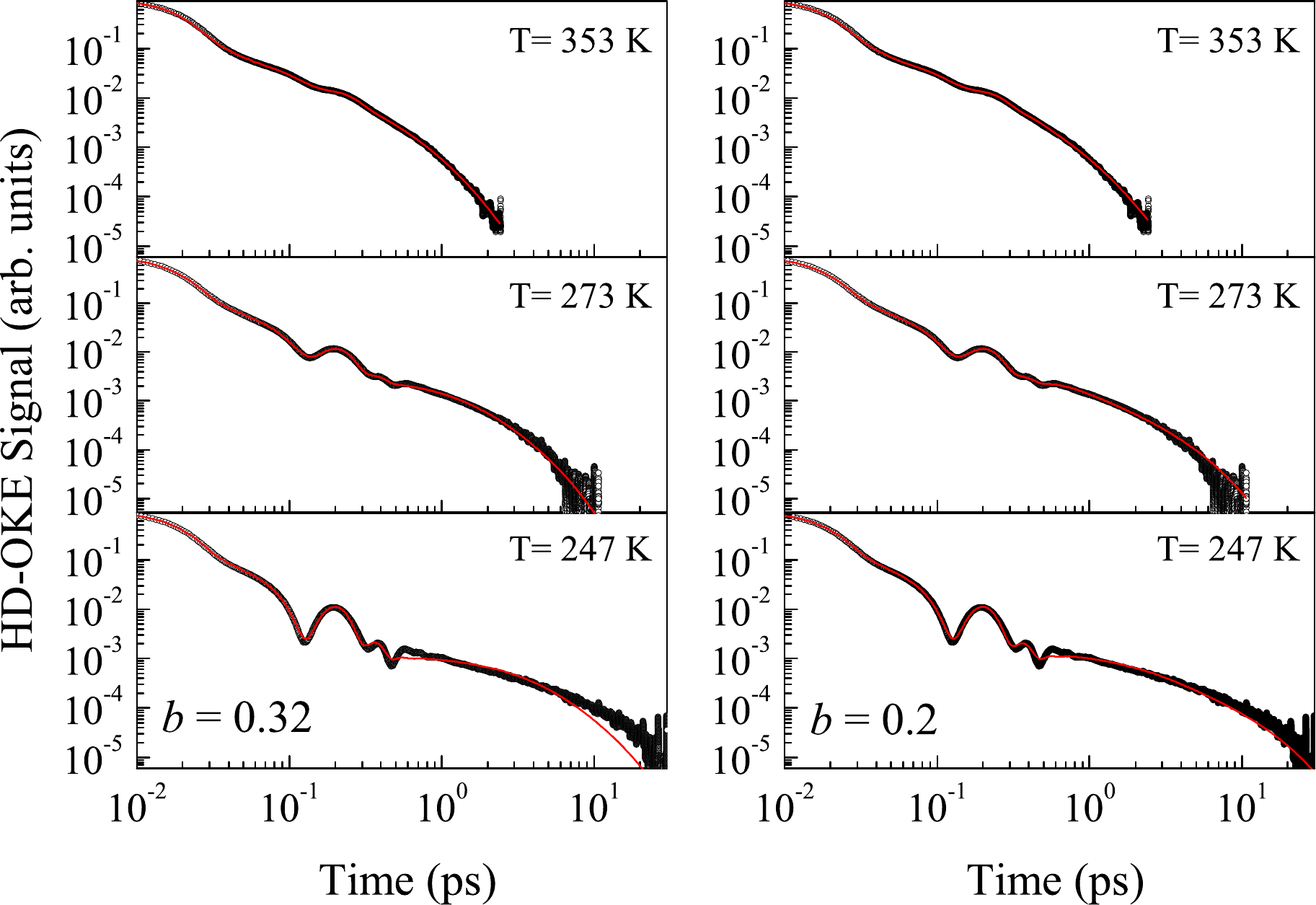}
\caption{Comparison of the experimental data at $247~K$, $273~K$, and $353~K$, and the MBO fits with cut-off proportionality factor $b=0.32$ and $b=0.2$}
\label{figMuka1}
\end{figure*}
In the liquid phase, we confirm the results already reported in ref. \cite{palese_96}, the model is able to reproducing the OKE signal in the whole temporal range, with the proper temperature behaviour of all the fitting parameters, but fails more and more as temperature decreases, see Fig.~\ref{figMuka1}. In summary, it is not able to accurately reproducing the fast dynamics, responsible for the complex structure of the vibrational bands and the stretched exponential character of the slow relaxation.

\subsubsection{MBO fitting details and results}

The fitting function is obtained by numerical convolution of the OKE response function, see Eqs.~\ref{signalfit} and Eq.~\ref{RespMBOtot}, with the instrumental function obtained as explained in Section~\ref{InstrFunct}. The best fit is achieved by the repeated variation, in a numerical loop, of the fitting parameters, until a nonlinear least-squares minimization is reached.

The fitting function is built on an inhomogeneous broadening presenting two water bands: at $180~cm^{-1}$ and $60~cm^{-1}$, which are usually assigned to stretching and bending of the hydrogen bond, respectively.  Palese et al. introduced other higher frequency bands in order to simulate their OKE data; according our analysis, frequencies higher then about $400~cm^{-1}$ are not necessary, once the instrumental response function is properly taken into account. 
In our MBO analysis, we used nine free fitting parameters: the parameters of the inhomogeneous distribution $A_1$, $\Omega_1$, $\Gamma_1$, $A_2$, $\Omega_2$, $\Gamma_2$, the width of the homogeneous distribution $\gamma$, and the amplitudes $a_1^2$ and $a_2^2$ appearing in Eq.~\ref{RespMBO}.

The cut-off frequency $\Omega_c$ in Eq. \ref{RespMBOtot} turns to be a very critical parameter in the fit, so we performed several fitting runs with different values of this coefficient. If the value fo the $b$ parameter in \ref{OmegaC} is locked in the range $0.1-0.15$, the fits are apparently fine but the inhomogeneous distribution parameters turn out to be non physical at high temperatures. In particular, we observed a narrowing of the modes with rising temperature. In fact, a value of $b$ in such a low range yields a large homogeneous width, necessary to fit the long times part of he data, with a consequent narrowing of the inhomogeneous modes required for the fitting of the oscillating part. With larger values of $b$ ($0.25-0.3$), the model fails to fit the long part of the data at the lowest temperatures, the simulated decay going to zero too fast. With intermediate values of $b$ ($0.15-0.2$), the fit fails both at short and long times; moreover, also in this case the inhomogeneous distribution has an unphysical temperature dependence.   
Finally, a fully free value of the $b$ free fitting parameter leads to an apparently better reproduction of the data in the whole temporal range. However, the  temperature trend of the $\gamma$ parameter is definitely unphysical: in fact, it remains constant at all temperatures, yielding a temperature independence of liquid viscosity.

\begin{figure}[htb]
\includegraphics[scale=0.5]{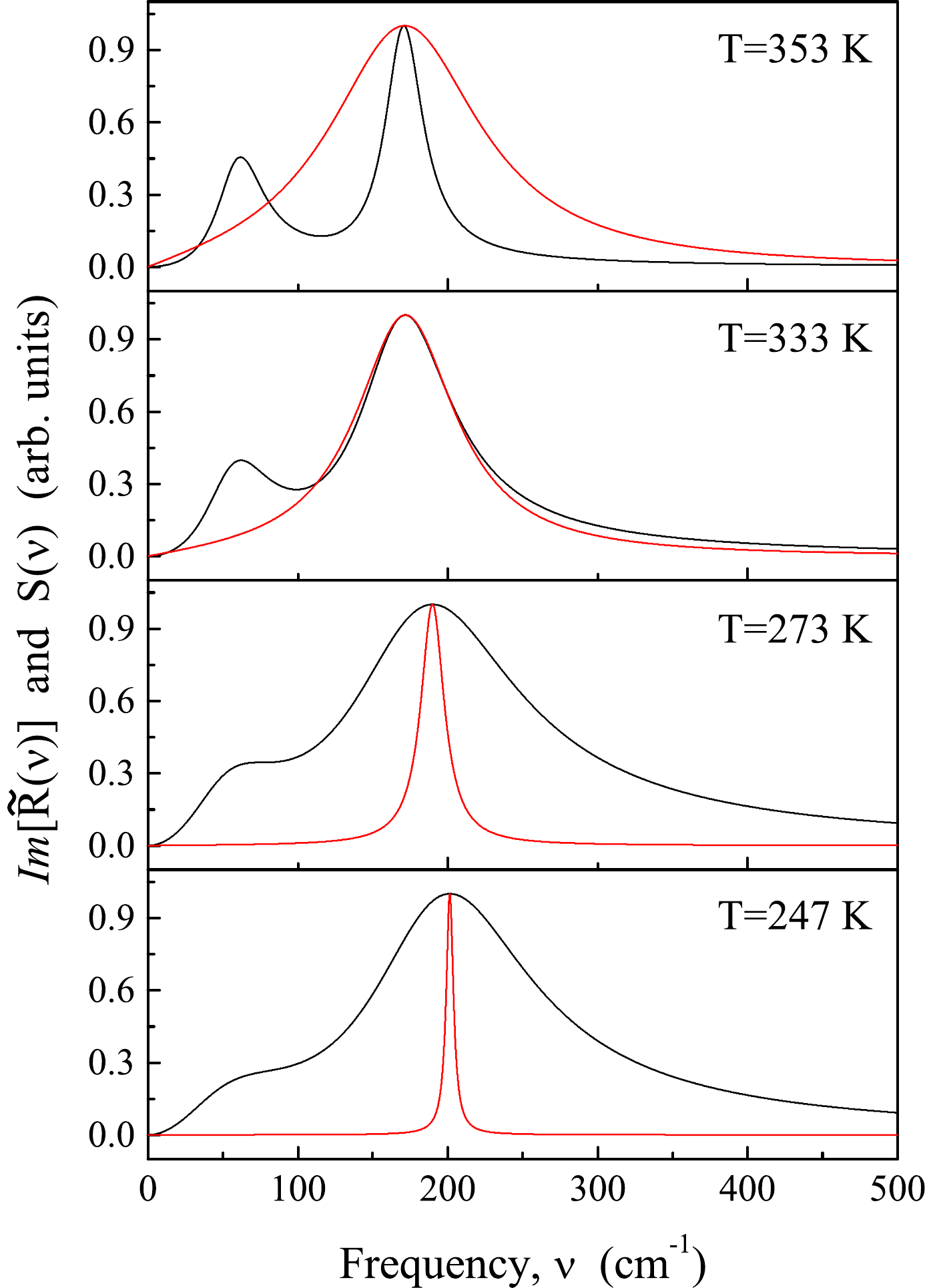}
\caption{We report the Fourier transform of the single oscillator response function (red line) and the inhomogeneous distribution function (black line), calculated using the parameters from the fitting section with $b=0.2$. The opposite temperature behaviours of the inhomogeneous and homogeneous distributions are evident in the spectral representation.}
\label{figMuka3}
\end{figure}

In Fig.~\ref{figMuka1} we report the HD-OKE data at three temperatures in the liquid and supercooled phases (247 K, 273 K, and 353 K) and the corresponding curves calculated with $b=0.32$ and $ =0.2$. As noted above, the agreement at high temperature is good, but it worsens at low temperatures.
%figure7
\begin{figure*}[t]
\includegraphics[scale=0.7]{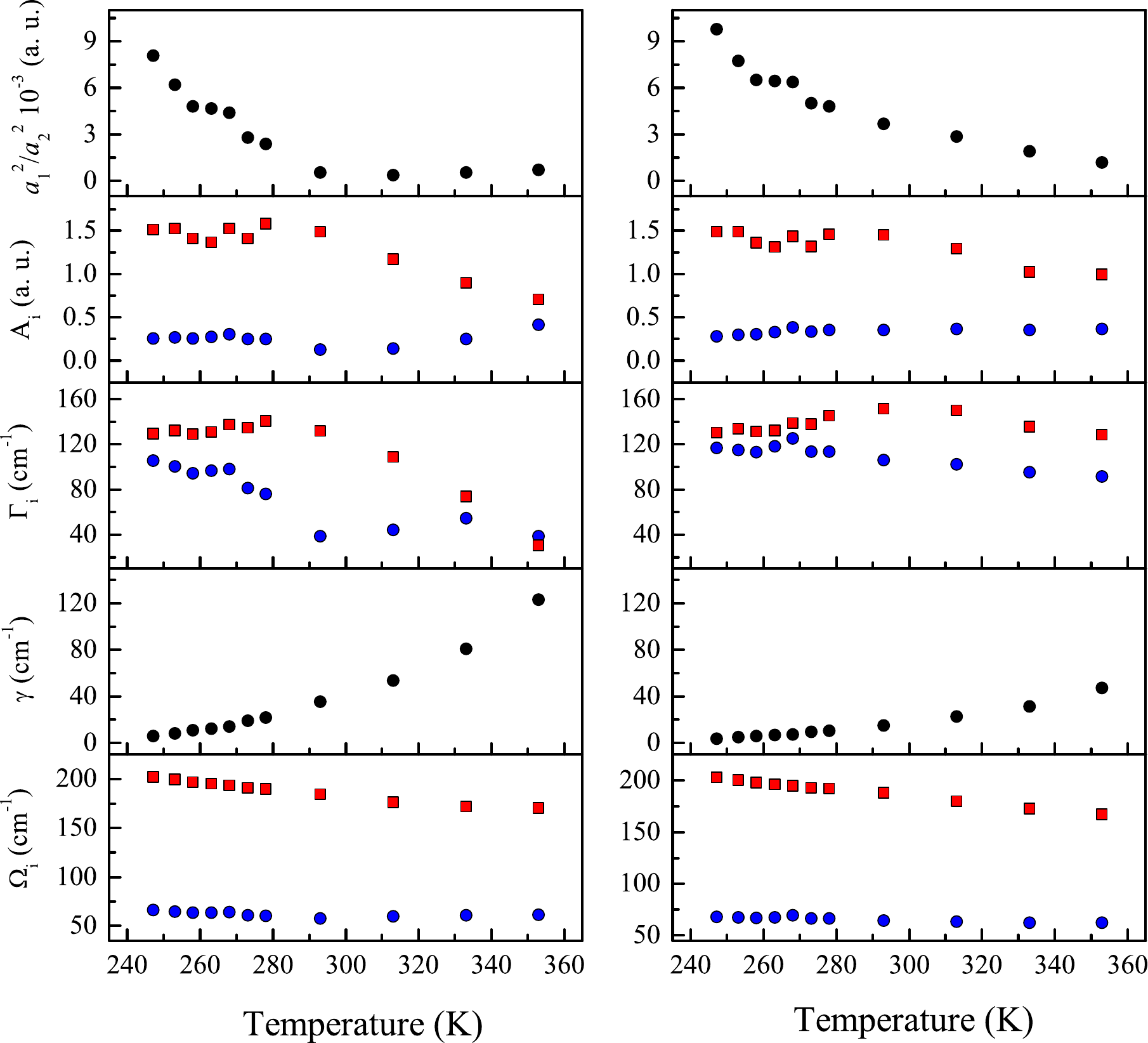}
\caption{Temperature dependence of the MBO fitting parameters for the fits with $b=0.2$ (left panels) and $b=0.32$ (right panels).}
\label{figMuka2}
\end{figure*}
The temperature dependence of the MBO fitting parameters for these two fitting series are shown in Fig.~\ref{figMuka2}. 
In Fig.~\ref{figMuka3} we report the imaginary part of the Fourier transform of the single oscillator response function, $\tilde{R}^{MBO}(\nu)$ and the inhomogeneous broadening  function $S(\nu)$. These are calculated adopting for the fitting parameters the values reported in the left panel of Fig.~\ref{figMuka2}, corresponding to $b=0.2$. The temperature dependence of the homogeneous broadening (black line) is the expected one, while the inhomogeneous distribution (red line) narrows as temperature increases, with a definitely non-physical behaviour.
%figure8

\subsection{Kubo's Discrete Random Jump Model}
Winkler et al.\cite{winkler_02} used a model based on the Kubo's Discrete Random Jump (KDRJ)~\cite{kubo_62,kubo_69} to fit the OKE data of liquid water. Their approach implies an a priori separation between the fast vibrational dynamics, described by the KDRJ model, and the slow relaxation, which is simulated by the time derivative of a stretched exponential function.   

The KDRJ model describes the dynamics of the liquid in terms of $Q(t)$ stochastic oscillators: $\ddot{Q}(t)-\frac{\dot{\omega}}{\omega}\dot{Q}(t)+\omega^2 Q(t)=0$; the $\omega(t)$ frequency is a stochastic variable randomly perturbed by $N$ independent two-state jump Markov processes (random-telegraph process), $\omega(t)=\Omega+\sum_{n=1}^N\omega_n(t)$. Each stochastic process is considered stationary, ie. $\langle\omega_n(t)\rangle=0$ and $\langle \omega(t)\rangle=\Omega$, and Markovian, with $\langle\omega_n(t)\omega_m(0)\rangle=\frac{\Delta^2}{N}exp(-\gamma\vert t \vert)$. In the latter expression, $\gamma$ is the rate of the random frequency modulation and $\Delta^2$ is the amplitude of the modulation. The total stochastic process is assumed Gaussian.

In the approach of Winkler et al.\cite{winkler_02}, the susceptibility tensor is linearly connected to the nuclear $Q$-modes, $\chi(t)\propto Q(t)$; thus, the OKE response function can be obtained from the time derivative of the $Q$-mode correlation functions. The contribution of a single $Q$-mode to the response function is expressed as: 
\begin{widetext}
\begin{equation}
R^{KDRJ}(\Omega,t)= \theta(t)\bigg[\frac{\gamma}{2}N\frac{a^2-1}{a}sinh\left(\frac{\gamma t}{2a}\right)cos(\Omega t)+ \psi(t)\Omega sin(\Omega t)\bigg] \psi(t)^{N-1} exp\left(-N\frac{\gamma t}{2}\right)
\label{Respkubo}
\end{equation}
\end{widetext}
with
\begin{equation}
\psi(t)=\left[cosh\left(\frac{\gamma t}{2a}\right)+a sinh\left(\frac{\gamma t}{2a}\right)\right] 
\end{equation}
where
\begin{equation}
a=\left(1-4\frac{\Delta^2}{\gamma^2}\right)^{-1/2}
\label{psia}
%\end{align}
\end{equation}

The Fourier transform of the KDRJ response function $Im [\tilde{R}^{KDRJ}(\nu)]$ provides the spectral representations of the involved dynamics. The resulting spectral profile is strongly dependent on the $\Delta/\gamma$ ratio. For $\Delta/\gamma\gg 1$, slow modulation limit, the line consists of $N+1$ lines with an overall Gaussian envelope profile whose full width half maximum is equal to $\Delta$. The lines are spectrally separated by the quantity $2\Delta/\sqrt{N}$, and each line is homogeneously broadened with width $\gamma$ due to the finite life time of the level itself. For $\Delta/\gamma\ll 1$, motional narrowing limit, the multiplet structure collapses in a single resonance. In this case the frequency jumps occurs on a time scale faster that the average vibrational period $2\pi/\Omega$. It is clear that, depending on the average frequency $\Omega$, on the $\Delta/\gamma$ ratio, and on the number of stochastic processes $N$, we can obtain complex line shapes, which describe the structured vibrational bands of liquids.

In their data analysis, Winkler et al. introduced an extra interaction between the Kubo oscillator and a further thermal bath. This interaction produces an inhomogeneous broadening defined by\cite{winkler_02}:
\begin{equation}
S_i(\Omega)=exp\left[\frac{-4ln(2)(\Omega-\Omega_i)^2}{\Gamma_i^2}\right]
\label{SOmegaK}
\end{equation}
and the oscillator response function becomes:
\begin{equation}
R_i^{KDRJ}(t) \propto \int_{0}^{\infty} \left[S_i(\Omega)R^{KRDJ}(\Omega,t)\right] d\Omega
\end{equation}
If more Kubo oscillators are involved in the dynamics,
\begin{equation}
R_n(t)=\Sigma_i R_i^{KDRJ}(t)+\theta(t)At^{\left(\beta-1\right)}\exp\left[-\left(\frac{t}{\tau_s}\right)^{\beta}\right]
\label{RespKDRJtot}
\end{equation}
$\tau_s$ being the structural relaxation time and $\beta$ is the stretching factor. The presence of stretched exponential decay in OKE data has been proved to show-up both in glass-former liquids\cite{torre_98} and supercooled water\cite{torre_04}.

Eq.\ref{RespKDRJtot} implies that the vibrational dynamics, described by the $Q$-modes, is uncoupled from the structural relaxation, described by the stretched exponential decay. If the time/energy scale of the structural relaxation could be considered well separated from that of other dynamics, the decoupling hypothes would be properly founded. In water both structural relaxation and vibrational dynamics take place on very similar time/energy scales\cite{torre_04,taschin_13}; the same is true for the H-bond dynamics\cite{fecko_03}. Any decoupling approximation then appears a rather unrealistic hypothesis. Apart from these fundamental criticisms, we tested the ability of the model to fit our OKE data in supercooled water. 

\subsubsection{KDRJ fitting details and results}

The fitting function was obtained as the convolution of the response function with the instrumental function. The OKE nuclear response function is simulated, following the analysis of Winkler et al. analysis, using two KDRJ oscillators with $N=3$. The free fitting parameter were: four parameters of each oscillators, $\Delta_i$, $\gamma_i$, $\Omega_i$, and $\Gamma_i$, with the constrain $\gamma_1=\gamma_2$, two amplitudes $a_i$, and the stretched exponential parameters, $A$, $\tau_s$, and $\beta$.
%figure9
\begin{figure}[htb]
\includegraphics[scale=0.6]{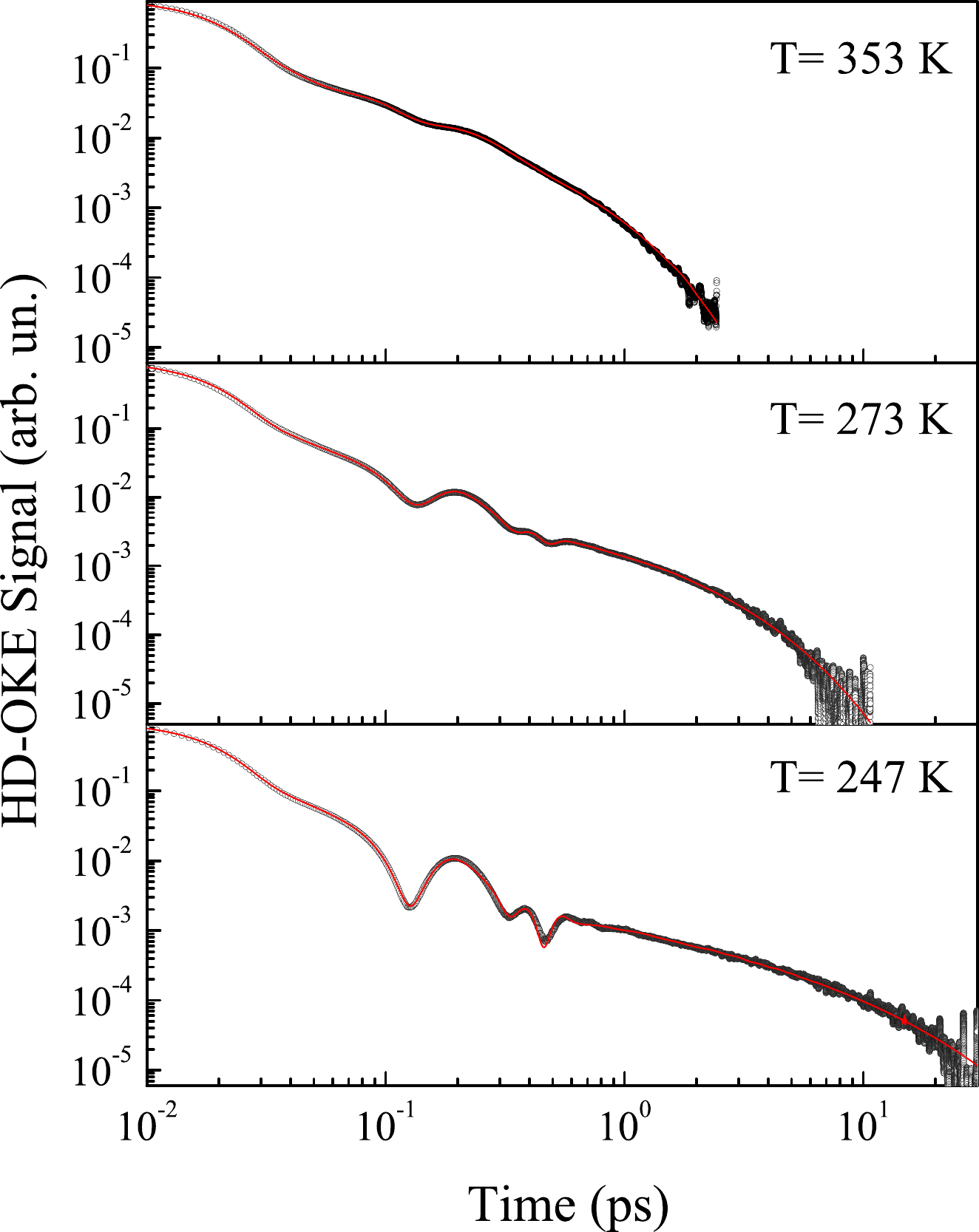}
\caption{Comparison between the experimental data and the KDRJ model fits at $247~K$, $273~K$, and $353~K$.}
\label{figKubo1}
\end{figure}
We report in Fig. \ref{figKubo1} the fit-data comparison for the three temperatures $247~K$, $273~K$, and $353~K$. The long time part of the signal is clearly very well described by the adopted stretched exponential function. The short time oscillating part is fairly well reproduced at high temperatures, as already found in ref.~\cite{winkler_02}. In the deeply supercooled phase this part is not perfectly reproduced (see the oscillations around $1~ps$) but the model is able to account for the growing structuring of the vibrational bands at low temperatures. 
%figure10
\begin{figure}[htb]
\includegraphics[scale=0.45]{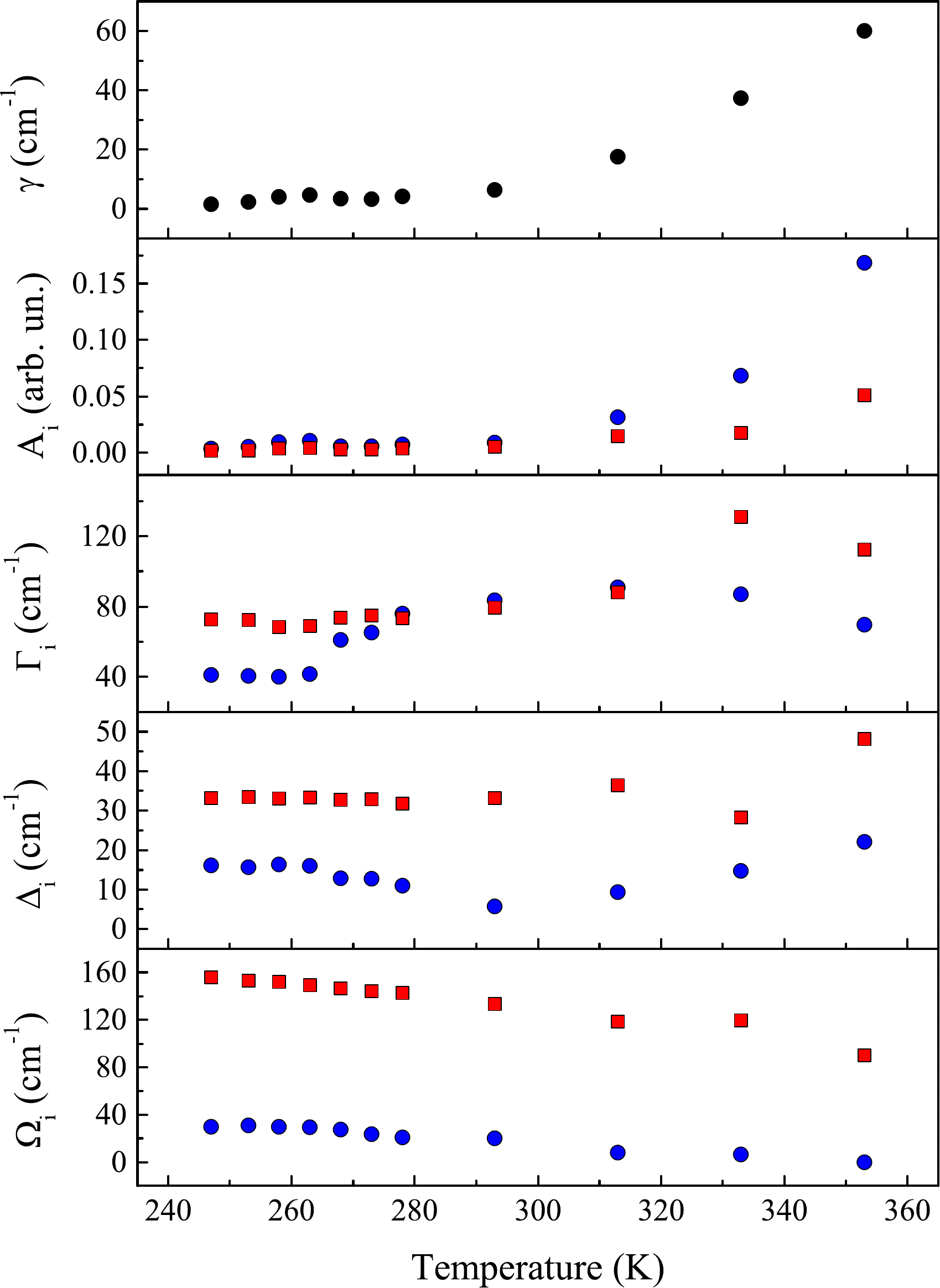}
\caption{Temperature behaviour of the best fit parameters for the two KDRJ oscillators.}
\label{figKubo2}
\end{figure}
In Fig.~\ref{figKubo2}, we report the values of the fitting parameters for all the temperatures.
%figure11
\begin{figure}[htb]
\includegraphics[scale=0.5]{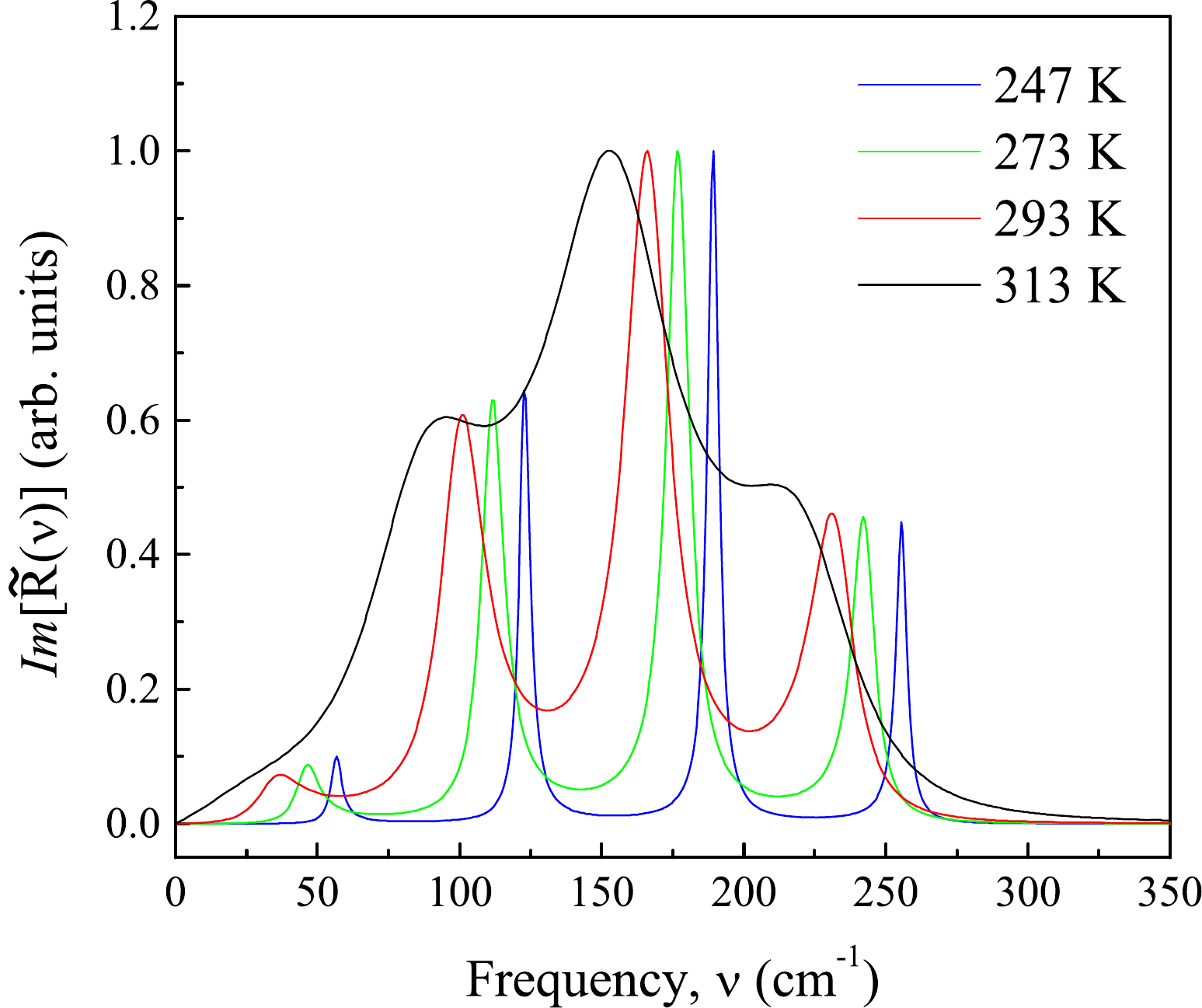}
\caption{Temperature behaviour of the spectral shape of the KDRJ oscillator centred around $180~cm^{-1}$, stretching band, neglecting the inhomogeneous broadening. The spectrum peaks could be addressed to different water clusters.}
\label{figKubo3}
\end{figure}

It is worth to look at the spectral shape of the KDRJ oscillator when $S_i(\Omega)=\delta(\Omega-\Omega_i)$, i.e. in the hypothesis of neglecting the inhomogeneous broadening due to the interaction with the thermal bath. In Fig.\ref{figKubo3} we show the imaginary part of the Fourier transform the single KDRJ oscillator response function, $Im[\tilde{R}^{KDRJ}(\nu)]$, at different temperatures. For each temperature the $\Omega$ parameter is fixed to the highest value obtained from the fitting procedure, see   
Fig.\ref{figKubo2}. The spectrum of this KDRJ oscillator represent the homogeneous vibrational components describing the stretching water band. Clearly $N+1=4$ resonances are present. 

Winkler et al. proposed an intriguing interpretation of these resonances for liquid water: they would correspond to the H-bond stretching frequencies of different molecular aggregates, from dimer units to the tetrahedral pentamer units. These clusters inter-convert each other through breaking and making of the H-bonds; this very fast process would result int the KDRJ frequency jumps. The frequency and damping of the KDRJ oscillator would be related to the H-bond stretching/bending vibrations and to their life time, respectively. The application of this picture to the OKE data analysis suggests that liquid water consists of mixture of four different clusters, having comparable and weakly temperature dependent concentrations. The most recent experimental investigations\cite{nilsson_12,taschin_13}, simulation studies\cite{overduin_12,kesselring_12} and theoretical models\cite{holten_12,holten_13,tanaka_13} do not support this scenario: the emerging picture is that water presents a bimodal local structuring (i.e. formation of two main molecular clusters); water molecules form local structures either tetrahedrally coordinated, named low-density forms, or close-packed, named high-density form. Moreover, the relative populations of these two alternative local structures are strongly temperature dependent.

\subsection{Schematic Mode-Coupling Model}

The Mode-Coupling Theories (MCT)\cite{goetze_09} are a generalization of the Mori and Zwanzig approach. The liquid dynamics is described by the memory-function equations, which define the equations of motion of the correlation functions of physical observables. The retardation effects are taken into account by the memory functions $K(t)$, that in the MCT are defined on the basis of the correlators. These theories represent a generalized hydrodynamic approach to the liquid dynamics where the physical observables are intrinsically mesoscopic.

In the schematic mode-coupling (SMC) model~\cite{goetze_92,goetze_00b,goetze_04,goetze_09} the main variable is the density, $\rho(t)$; hte other observables, $Q_i(t)$, are linked to the density.  
 
The time evolution of the correlation functions of these physical observables is given by the memory-function equations. They are formulated as~\cite{goetze_92,goetze_00b,goetze_04,goetze_09}:
\begin{equation}
\ddot{\Phi}_m(t)+\eta_m\dot{\Phi}_m(t)+{\Omega_m}^2\Phi_m(t)+
	\int K(t-t')\dot{\Phi}_m(t')dt'=0
	\label{mastereq}
\end{equation}
with the memory function written as
\begin{equation}
	K(t)= v_1 \Phi_{m}(t)+v_2 \Phi^2_{m}(t)
	\label{masterememo}
\end{equation}
The SMC model defines the memory by a series expansion (up to the second term) of the \textit{master correlator} itself $\Phi_{m}$, thus providing a closed form for the integro-differential equation \ref{mastereq}. The SMC model identifies the master correlator as the density correlator $\Phi_m\propto\langle\rho(t)\rho(0)\rangle$; the quadratic term in eq. \ref{masterememo} corresponds to the minimum order of the series expansion able of reproducing the slowing down behaviour of the structural relaxation.
The dynamics of any other observable, $Q_i$, linked to the time dependent density (e.g. to the local inter-molecular dynamics) can be described by a similar differential equation~\cite{bosse_87b}:
\begin{equation}
\ddot{\Phi}_i(t)+\eta_i\dot{\Phi}_i(t)+{\Omega_i}^2\Phi_i(t)+
	\int m_i(t-t')\dot{\Phi}_i(t')dt'=0
	\label{slaveeq}
\end{equation}
In \ref{slaveeq} the memory is given by
\begin{equation}
	m_i(t)= v_i^s\Phi_m(t)\Phi_i(t)
\label{slavememo}
\end{equation}
 $\Phi_i(t)\propto\langle Q_i(t)Q_i(0)\rangle$ being the \textit{slave correlator}. The coupling between the slave and master dynamics is assured by the product of the slave and master correlators in the memory kernel \ref{slavememo}.

Equations \ref{mastereq}, \ref{masterememo}, \ref{slaveeq}, and \ref{slavememo} are a closed set that can be solved numerically, as analytic solutions exist only in a restricted number of cases\cite{goetze_09}.

MCT is essenzially a hydrodynamic model: than is hard to attribute a precise microscopic (at the molecular level) interpretation of the involved physical quantities; the $Q_i$ variables can be interpreted as key parameters influencing the liquid susceptibility. Their dynamics, described by $\Phi_i$ correlators, allow the calculation of the OKE nuclear response function. The experimental response can be expressed as the time derivative of the sum of these slave correlators:
\begin{equation}
R_n(t)\propto -\theta(t)\frac{\partial}{\partial t}\sum_i a_i \Phi_i(t).
\label{mct-response}
\end{equation}
In other words, equations \ref{slaveeq}, \ref{slavememo} and \ref{mct-response} correspond to decomposing the electronic susceptibility correlator, $\Phi_{\chi\chi}$, into the sum of $\Phi_i(t)$ correlators. Each of these correlator describes an ``average collective mode", whose dynamics is described by the SMC equations. The vibrational and relaxation properties and the coupling of different observables are present into the SMC equations by definition. In this respect, the SMC equations represent a robust physic model able of describing complex dynamics including damped vibrations and structural relaxation, as well as their coupling. Differently from other approaches, SMC does not require any decoupling or dynamic separation between the fast/vibrational dynamics and the slow/relaxation phenomena.  

\subsubsection{SMC fitting details and results}

We solved the SMC equations numerically, taking the frequencies, friction and coupling coefficients as parameters to be adjusted in order to reproduce the HD-OKE response. We adopted a step-by-step second order Runge–Kutta algorithm to solved numerically the integro-differential equations Eq.\ref{mastereq} and Eq.\ref{slaveeq}.  Once the time dependent function of the master correlator is known, it can be used for calculating those of the slave correlators and then that of the OKE signal. The parameters of the model are: the master equation parameters $\eta_m$, $\Omega_m$, $v_1$ and $v_2$, the slave equations parameters $\eta_i$, $\Omega_i$, $v_i^s$ with $i=1,2,3$ and the three amplitudes $a_i$ in eq. \ref{mct-response}. Of course, the result of the fit depends on the number of slave correlators included: we considered the cases corresponding to one, two, and three correlators. We performed a preliminary series of fits to obtain a qualitative estimate of the temperature dependence of the parameters. On that basis, we chose, in agreement with what already done in similar analyses reported in literature \citep{alba_95,wuttke_00,goetze_00b,krakoviack_02,wiebel_02,goetze_04,ricci_04}, to force some of the parameter either to assume fixed values or to follow pre-established temperature trends. 

%Figure12
\begin{figure}[htb]
\includegraphics[scale=0.5]{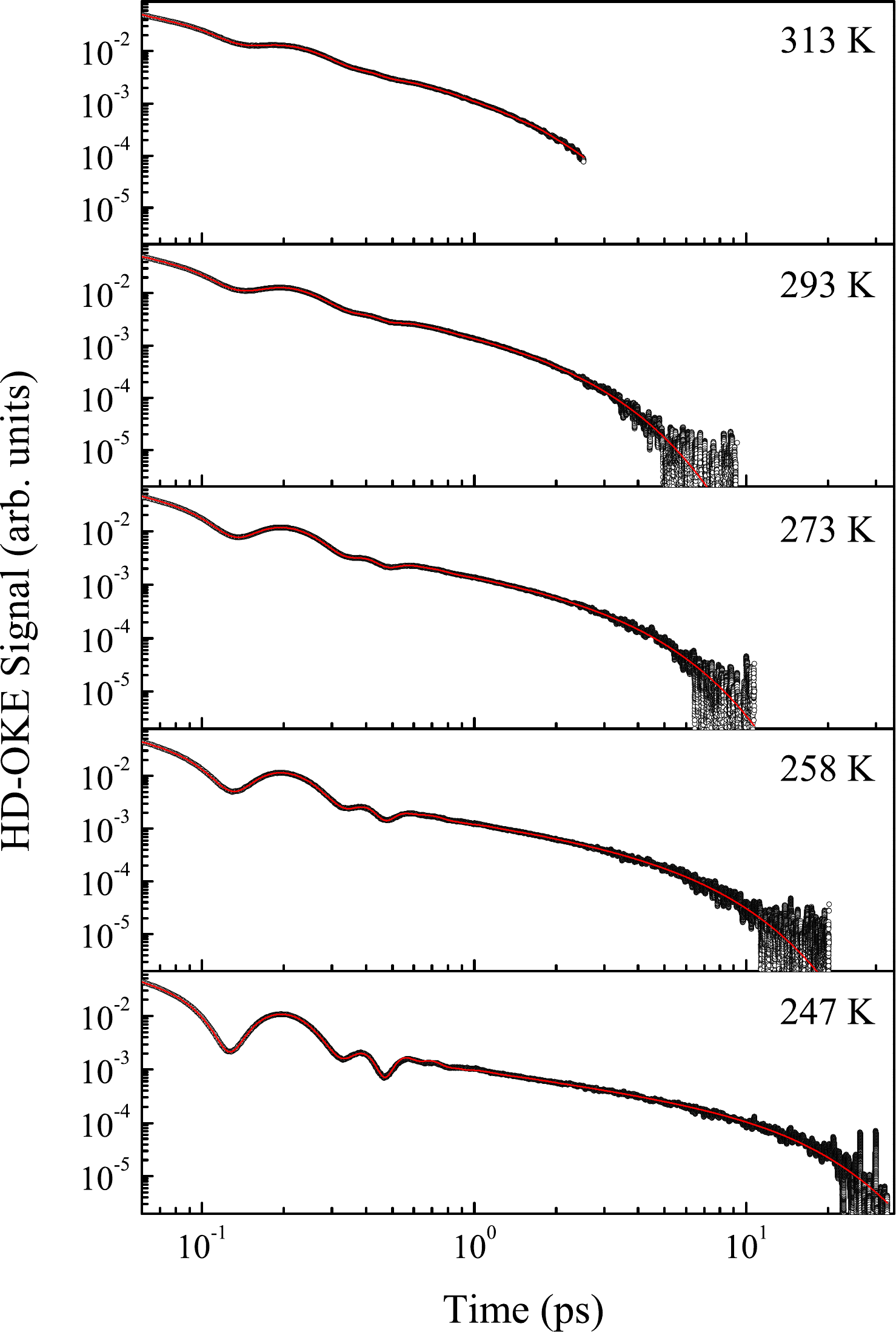}
\caption{SMC fits (red line) of the heterodyne-detected optical Kerr effect data (circle) at different temperatures in a log-log plot. The SMC model very well reproduces the complex vibrational dynamics taking place in the sub-picosecond time scale and the slow relaxation decays. Also at intermediate delay times, where the vibrational dynamics merge into the relaxation processes, the SMC equations correctly descibe the decay curve.}         
\label{SMC}
\end{figure}

We found that the OKE data can be described in all the temperature range with three slave correlators at most. In particular, the weight of the highest frequency contribution decreases monotonically as the temperature increases, while it becomes negligible at the two highest temperatures, where only two slave correlators are sufficient to reproduce the data. 

In Fig.~\ref{SMC} we show some of the measured HD-OKE signals with the fits obtained using the SMC model. The model reproduces correctly the experimental data over the whole time range at all temperatures. Most remarkably, and differently from other fitting models, this result is achieved without imposing any decoupling of vibrational and relaxation dynamics. 

%Figure13
\begin{figure}[htb]
\includegraphics[scale=0.4]{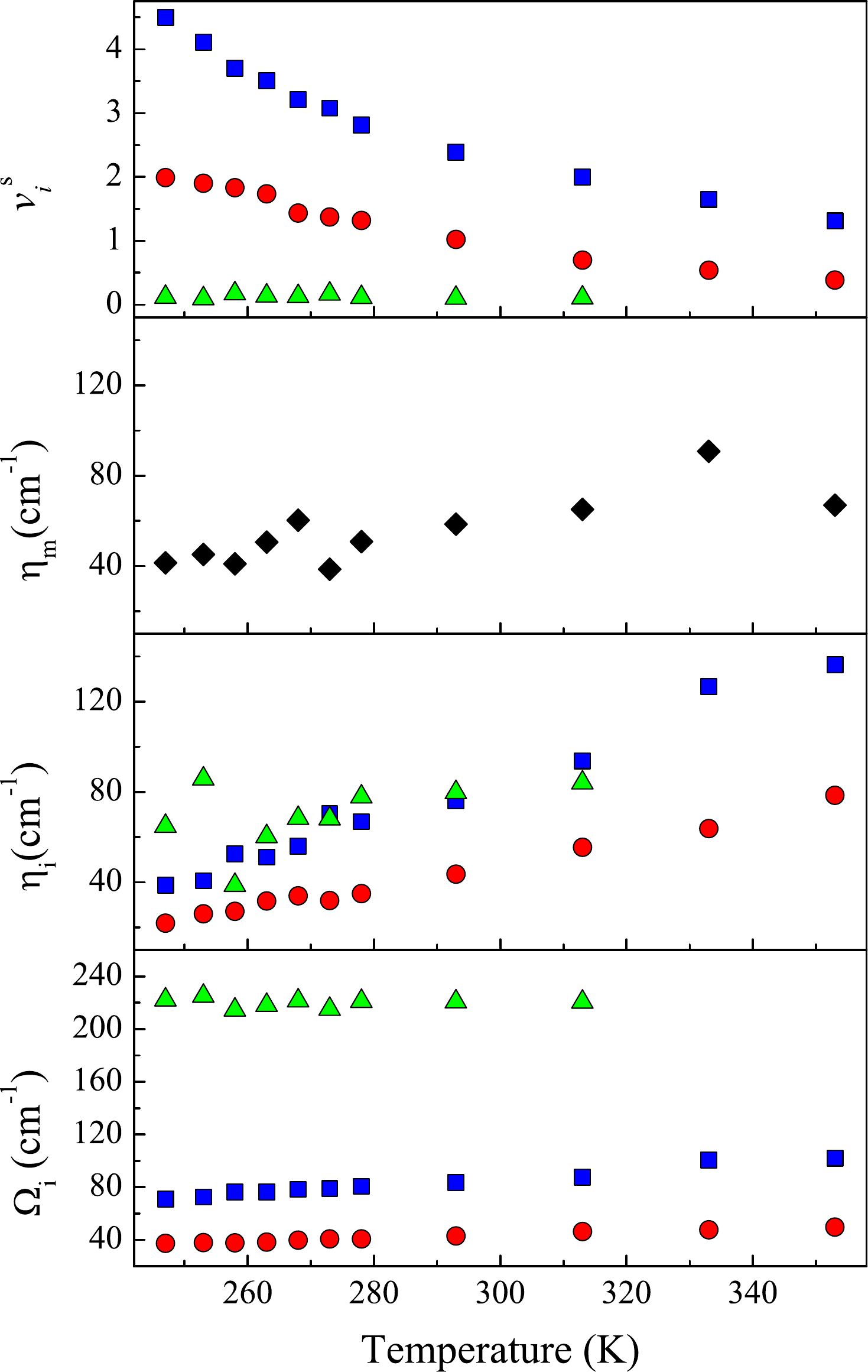}
\caption{Temperature behaviour of the fitting parameters of the SMC model, the slave frequencies $\Omega_i$, the friction parameters $\eta_m$ and $\eta_i$ and the three vertices $v^s_i$. Circles, squares, and triangles refer to $\Phi_i$ slave correlators, with $i=1$, $2$, and $3$, respectively; diamonds represent the master correlator.}
\label{MCTpar}
\end{figure}

The best values of the frequency $\Omega_m$ and and the vertex $v_1$ of the master oscillator were almost constant in the whole temperature range; then, they were locked to $66~cm^{-1}$ and $0.33$, respectively. The second vertex $v_2$ was, instead, resulted to increase almost linearly with decreasing temperature: it was forced to obey the linear dependence $v_2=6-0.014T$. Finally, we left free the friction $\eta_m$ and the remaining parameters of slave oscillators. In Fig.\ref{MCTpar} we show the temperature dependence of the slave frequencies $\Omega_i$, of the friction parameters $\eta_m$ and $\eta_i$ and of the three vertices $v^s_i$.

%Figure14
\begin{figure}[htb]
\includegraphics[scale=0.55]{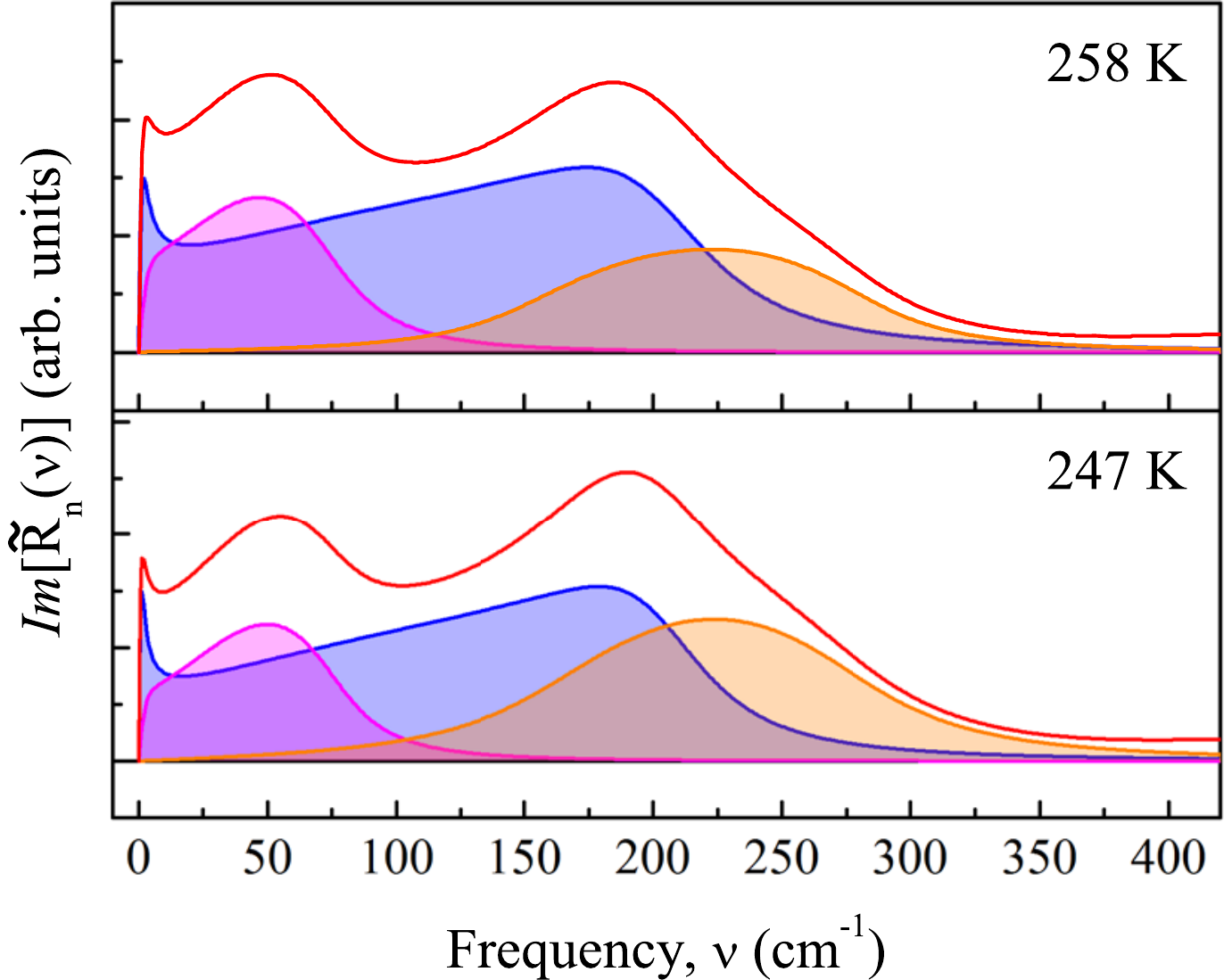}
\caption{ The Fourier transform of the SMC fit response function, $Im[\tilde{R}_n(\nu)]$, is reported (red line). The three correlators $Im[\tilde{\Phi}_{1,2,3}(\nu)]$, are also shown (magenta-blue-orange areas). The simulation of HD-OKE data based on the SMC model requires two modes (blue and orange shaded areas) to fit the high frequency band. The characteristics of these two modes are clearly different in terms of spectral shape and temperature dependence.}
\label{MCT-FFT}
\end{figure}

In Fig.\ref{MCT-FFT} we report the imaginary part of the Fourier transform of the SMC simulated OKE responses obtained by the best fit of the experimental data at two temperatures. The contributions of the three slave correlators are reported in the figure as magenta-blue-orange lines. The simulation of HD-OKE data by SMC model requires two vibrational modes to fit the intermolecular stretching band of water (blue and orange lines). The characteristics of these two modes are clearly different in terms of spectrum shape and temperature dependence. As discussed in a previous paper\cite{taschin_13}, these two modes can be  associated with two fluctuating water species with different local structures; a low-density form characterized by a tetrahedral network and a high-density form characterized by closely packed aggregates with lower coordination and high network distortions.

\subsection{Final considerations}
Two points of general relevance for the OKE investigation of the dynamical properties of molecular liquids come out from the above discussion: i) the analysis of the time domain OKE response and, most of all, of its spectral representation, is critically dependent on the accurate determination of the instrumental function. Only an absolutely faithful determination of the latter can provide a reproducible response function, susceptible to reliable detailed analysis; ii) extending the investigation to low temperatures is an essential requirement: different models can provide equivalent results at high temperature, while diverging in their heuristic power when confronted to low temperature experiments. In fact, only at low temperature non-exponential behaviors of the OKE signal decay can show-up, and marked structuring of the oscillating part can grow-in.
Besides these general aspects, the main goal of our work has been that of analyzing the ability of different theoretical models to reproduce femtosecond HD-OKE data of very high quality and accuracy. The three models that we considered derive from significantly different approaches. In particular, the KDRJ model adopted by Winkler et al.\cite{winkler_02} differs significantly from the others as it assumes that the oscillatory and diffusive dynamics of liquid water can be separated a priori. In fact, the long time relaxation is described as an exponential decay and is subtracted from the HD-OKE time domain data, thus isolating the vibrational component. For the short time dynamics, the authors adopt an essentially molecular picture, based a Kubo treatment of the linewidth, which involves three intermolecular vibrational frequencies as stochastic variables. Ref.\cite{winkler_02} takes into account only room temperature data: we show here that extending the analysis to low temperatures, well below the thermodynamic melting point, provides a much more stringent test of the theory. In fact, at low temperature not only the long time relaxation has to be described as a stretched exponential decay, but also the agreement for the oscillatory part of the OKE response, very good at room temperature, is definitely less satisfactory in the supercooled regime. In any case, the KDRJ approach accounts fairly well for the growing structuring of the oscillatory pattern at low temperatures.
The other two models considered get rid of the imposed separation of the diffusive contribution from the overall dynamics, a separation that appears hardly justified in view of the similar time scales of the structural relaxation and of the intermolecular vibrations of liquid water. Similarly to KDRJ, the Brownian oscillator (MBO) approach, employed for water by Palese et al.\cite{palese_96}, is based on an almost microscopic picture, whose dynamical variables consist of a collection of averaged local intermolecular modes. Under-damped oscillators of relatively high frequency account for the oscillatory part of the response, while the structural relaxation contributions originate from the superposition of over-damped oscillators described by eqs.\ref{SOmega} and \ref{RespMBOtot}. We found that the most critical parameter is the cut-off frequency $\Omega_c$ in eq.\ref{RespMBOtot}: it is this low frequency limit that inhibits a good fit of the non-exponential decay present in the low temperature OKE data. Nevertheless this limit is imposed by the physical restriction that the oscillator period cannot be longer than the structural rearrangement.
The mode coupling (SMC) treatment is based on a continuum picture of the liquid, and describes its dynamics on the basis of time correlation functions of physical observables. The key points are the non-linear form of the master memory function, eq.\ref{masterememo}, in the equation of motion of the density correlator, eq.\ref{mastereq}, and the inclusion of slave correlators coupled to the density correlator by the slave memory kernels, eq.\ref{slavememo}. We found that the SMC model provides the most flexible set of equations and allows a very good fit at all temperatures of the OKE data in the entire experimental time window. This approach is essentially hydrodynamic, hence no immediate link can be made between those correlators and the inter- and intra-molecular modes typical of a molecular-scale description of the dynamics. Only in particular conditions a link with the specific molecular features can be out-lined\cite{goetze_09}. In this sense, it appears mostly suitable to the investigation of pre-transitional and critical phenomena. The most interesting feature of the SMC analysis of the experimental data is that, using a rigorous physical approach and avoiding questionable assumptions, it allows disentangling dynamical contributions characterized by peculiar temperature (and, possibly, pressure) dependence. 

\section*{Acknowledgments}

This work was supported by Regione Toscana,  prog. POR-CRO-FSE-UNIFI-26, by Ente Cassa di Risparmio Firenze, prog. 2012-0584 and by MIUR, prog. PRIN-2010ERFKXL-004.  We acknowledge M. De Pas, A. Montori and M. Giuntini for providing their continuous assistance in the electronic set-ups ; R. Ballerini and A. Hajeb for the mechanical realizations.

%\bibliography{MyPaper,water}

\end{document}